\documentclass[11pt,a4paper]{article}
\usepackage{amsmath,amssymb}
\usepackage{epsfig,graphicx}

\newcommand{\reali}{\hbox{\rm I\hskip-2pt\bf R}}
\newcommand{\complessi}{\hbox{\rm I\hskip-5.9pt\bf C}}

\begin{document}

\title{\bf Rigorous Definition of
Quantum Field Operators in Noncommutative Quantum Field Theory\footnote{Talk
given at the 15th International Seminar on High Energy Physics QUARKS'2008,
Sergiev Posad, Russia, May 23-29, 2008 (to appear in the Proceedings).}}

\author{M.~N.~Mnatsakanova$^{a}$
and Yu.~S.~Vernov$^{b}$
\\
$^a$ \small{\em Institute of Nuclear Physics, Moscow State University} \\
\small{\em  119992, Vorobyevy Gory, Moscow, Russia } \\
$^b$ \small{\em Institute for Nuclear Research, Russian Academy of Sciences} \\
\small{\em 60-the October Anniversary prospect 7a, Moscow, 117312, Russia } }
\date{}
\maketitle{}

\begin{abstract}
The space, on which quantum field operators are given, is constructed in any
theory, in which the usual product between  test functions is substituted by
the $\star$-product (the Moyal-type product). The important example of such a
theory is noncommutative quantum field theory (NC QFT). This construction is
the key point in the derivation of the Wightman reconstruction theorem.
\end{abstract}

\section{{Introduction}}
Quantum field theory (QFT) as a mathematically consistent theory was
formulated in the framework of the axiomatic approach in the works of
Wightman, Jost, Bogoliubov, Haag and others  (\cite{SW} - \cite{Haag}).

Within the framework of this theory on the  basis of  most general
principles such as Poincar\'{e} invariance, local commutativity
and spectrality, a number of fundamental physical results, for
example, the CPT-theorem and the spin-statistics theorem  were
proven \cite{SW} - \cite{BLOT}.

Noncommutative quantum field theory (NC QFT) being one of the
generalizations of standard QFT has been intensively developed
during the past years (for reviews, see \cite{DN, Sz}). The idea
of such a generalization of QFT ascends to Heisenberg and  was
initially developed in Snyder's work \cite{Snyder}. The present
development in this direction is connected with the construction
of noncommutative geometry \cite{Connes} and new physical
arguments in favor of such a generalization of QFT \cite{DFR}.
Essential interest in NC QFT is also due to the fact that in some
cases it is a low-energy limit of string theory \cite{SeWi}.

The simplest and at the same time the most studied version of
noncommutative theory is based on the following Heisenberg-like
commutation relations between coordinates:
\begin{equation}\label{cr}
[\hat{x}_\mu,\hat{x}_\nu]=i\,\theta_{\mu\nu},
\end{equation}
where $\theta_{\mu\nu}$ is a constant antisymmetric matrix.

The relation (\ref{cr}) breaks the Lorentz invariance of the theory, while the
symmetry under the $SO \, (1,1) \otimes SO \, (2)$ subgroup of the Lorentz
group survives  \cite{AB}.

NC QFT can be formulated also in commutative space by formally replacing the
usual product of operators by the star (Moyal-type) product:
\begin{equation} \label{mprodx}
\varphi (x) \star\varphi (x) = \exp{\left ({\frac{i}{2} \,
\theta^{\mu\nu} \, \frac{ {\partial}}{\partial x^{\mu}} \, \frac{
{\partial}}{\partial y^{\nu}}} \right)} \,\varphi (x) \varphi
(y)|_{x=y}.
\end{equation}
This product of operators can be extended to the corresponding product of
operators in different points:
\begin{equation} \label{vfsprod}
\varphi (x_1) \star \cdots \star \varphi (x_n) = \prod_{a< b\leq
n} \,\exp{\left ({\frac{i}{2} \, \theta^{\mu\nu} \, \frac{
{\partial}}{\partial x^{\mu}_a} \, \frac{ {\partial}}{\partial
x^{\nu}_b}} \right)} \,\varphi (x_1)  \ldots  \varphi (x_n).
\end{equation}

Wightman approach in NC QFT was formulated in \cite{AGM} and
\cite{CMNTV}.  For scalar fields the CPT theorem and the
spin-statistics theorem were proven in the case $\theta^{0\nu} =
0$. In \cite{VM05} - \cite{CMTV} it was shown that the main
classical axiomatic results are valid or have analog in NC QFT at
least if time commutes with spatial variables, i.e. $\theta^{0\nu}
= 0$.

In \cite{AGM} it was proposed that Wightman functions in the
noncommutative case can be written down in the standard form
\begin{equation} \label{wf} W\,(x_1,  \ldots, x_n) = \langle \, \Psi_0, \varphi (x_1)
 \ldots \varphi (x_n) \, \Psi_0 \, \rangle,
\end{equation}
where $\Psi_0$ is the vacuum state. However, unlike the
commutative case, these Wightman functions are only $SO \, (1,1)
\otimes SO \, (2)$ invariant. In fact in \cite{AGM} the CPT
theorem has been proven in the commutative theory, where Lorents
invariance is broken up to $SO \, (1,1) \otimes SO \, (2)$
symmetry, as in the noncommutative theory it is necessary to use
the $\star$-product  at least in coinciding points.

In \cite{CMNTV} it was proposed that in the noncommutative case
the usual product of operators in the Wightman functions has to be
replaced by the Moyal-type product (\ref{vfsprod}) both in
coinciding and different points. Such a product of operators
reflects the natural physical assumption, that noncommutativity
should change the product of operators not only in coinciding
points, but also  in different ones.

Actually it seems very natural to use in different points not  the
$\star$-product itself, but the following generalization of the
$\star$-product in coinciding points:
\begin{equation} \label{vfprod}
\varphi\,(x)  \star \varphi\,(x)  \rightarrow \xi\,(x - y)\, \varphi\,(x)
\star \varphi\,(y), \qquad \xi\,(0) = 1,
\end{equation}
where $\xi\,(x - y)$ is some function rapidly falling if ${|x - y|}^2 \gg
\theta$. For example, $\xi\,(x - y)$ can be arbitrary continuous function,
satisfying the inequality:
\begin{equation} \label{exp}
\left| \xi\,(x - y)\right| \leq  C\, \exp {\left( - {{|x - y|}^2\over \theta
}\right)},
\end{equation}
where $C$ is a some positive number,
$$
\theta \equiv \max\limits_{\mu, \nu}|\theta^{\mu\nu}|, \quad |x -
y| \equiv \max\limits_{i = 0, 1, 2, 3}|x_{i} - y_{i}|.
$$
In \cite{VM05} it was shown that in the derivation of axiomatic
results, the concrete type of product of operators in various
points is insignificant.

The Wightman functions can be generally written down as follows
\cite{VM05}:
\begin{equation} \label{wfg}
 W\,(x_1,  \ldots, x_n) = \langle \Psi_0, \varphi \,(x_1)\,
 \tilde \star \,  \cdots   \tilde \star \, \varphi\,(x_n) \Psi_0 \rangle.
\end{equation}
The meaning of $\tilde\star$ depends on the considered case. In
particular,
\begin{equation}\label{forallxy}
\varphi \, (x) \tilde \star \varphi \, (y) = \varphi\,(x) \star
\varphi\,(y),
\end{equation}
\begin{equation}\label{formodxy}
\varphi\,(x)   \tilde \star \varphi\,(y) = \xi\,(x - y)\,
\varphi\,(x) \star \varphi\,(y),
\end{equation}
\begin{eqnarray}\label{forxy}
\varphi\,(x) \tilde \star \varphi\,(y) = \varphi\,(x)
\varphi\,(y), \;
x \neq y;  \nonumber\\
\varphi\,(x) \tilde \star \varphi\,(x) = \varphi\,(x) \star
\varphi\,(x)
\end{eqnarray}
It will be shown below that eqs. (\ref{forxy}) have to be substituted by their
regularization.

Let us stress that actually the field operator given at a point
cannot be a well-defined operator \cite{BLOT}. Well-defined
operator is a smoothed operator:
\begin{equation} \label{vff}
\varphi_f \equiv\int \,\varphi\,(x)  \,f\,(x) \, d \, x,
\end{equation}
where $f\,(x)$ is a test function. Correspondingly Wightman functions are
generalized functions in the space of test functions. In QFT the standard
assumption is that all $f\,(x)$ are test functions of tempered distributions.
On the contrary, in the NC QFT the corresponding generalized functions can not
be tempered distributions as the $\star$-product contains infinite number of
derivatives. As is well known (see, for example, \cite{SW}) that  there could
be only a finite number of derivatives in any tempered distribution.

The formal expression (\ref{wfg}) actually means that the scalar
product of the vectors $\Phi_{k} = \varphi_{f_k}\,  \cdots
\,\varphi_{f_1} \, \Psi_0$ and $\Psi_n = \varphi_{f_{k + 1}}\,
\cdots \,\varphi_{f_n} \, \Psi_0$ is the following:
$$
\langle \, \Phi_{k}, \Psi_n \, \, \rangle =
$$
$$
\int\, W\,(x_1,  \ldots, x_n) \,
\overline{f_1\,(x_1)}\,\tilde{\star}  \cdots \tilde{\star}
\overline{f_k\,(x_k)} \tilde{\star} f_{k + 1}\,(x_{k +
1})\tilde{\star}  \cdots \tilde{\star} \,{f_n\,(x_n)}
$$
\begin{equation} \label{scprod}
d\,x_1 \ldots d\,x_n,\quad W\,(x_1,  \ldots, x_n) \,
=\langle\Psi_0,\varphi(x_1)\cdots\varphi(x_n)\Psi_0\rangle.
\end{equation}
In paper  \cite{CMTV7} it was shown that the series
\begin{equation} \label{fsprodxy}
f\,(x) \star f\,(y) = \exp{\left ({\frac{i}{2} \, \theta^{\mu\nu}
\, \frac{ {\partial}}{\partial x^{\mu}} \, \frac{
{\partial}}{\partial y^{\nu}}} \right)} \,f\,(x)  f\,(y)
\end{equation}
converges if $f\,(x) \in S^\beta, \; \beta <  1 / 2$, $S^\beta$ is
a Gel'fand-Shilov space \cite{GSh}. The similar result was
obtained also in \cite{Sol07}.

\section{Rigorous Definition of Quantum Field Operators in NC QFT }

Let us define rigorously quantum field operator $\varphi_f$. To
this end we construct a closed and nondegenerate space $J$ such
that operators $\varphi_f$ be well defined on dense domain of
$J$.

As in the commutative case every vector of  $J$ can be
approximated with arbitrary accuracy by the vectors of the type
$$
\varphi_{f_1}\, \cdots \,\varphi_{f_n}\,\Psi_{0},
$$
where $\Psi_{0}$ is a vacuum vector. In other words the vacuum
vector  $\Psi_{0}$ is cyclic.

The difference of noncommutative case from the commutative one is that the
action of the operator $\varphi_f$ is defined by the $\star$-product.

Construction of space $J$ we begin with the introduction of set
$M$ of breaking sequences of the following kind
\begin{equation}\label{ser}
g = \{g_{0}, g_{1}, \ldots g_{k} \},
\end{equation}
where $g_{0} \in \complessi, \, g_{1} = g_{1}^{1} \, (x_{1}), x _ {1} \in
{\reali} ^4, g_{i} = g_{i}^{1} \, (x_{1}) \, \tilde{\star} \cdots
\tilde{\star} g_{i}^{i} \, (x_i), \, x_j \in {\reali} ^4, 1 \leq j \leq i$,
particularly $\tilde{\star}$ is determined by the formulae (\ref{forallxy}) -
(\ref{forxy}), $k$ depends on $g$. Addition and multiplication by complex
numbers of the above mentioned sequences are defined component by component.
It is obvious, that $C \, g \in M$, if $g \in M$.

The every possible finite sums of the sequences belonging $M$ form
space $J'_{0}$ on which action of the operator $\varphi_f, \, f =
f\,(x), \; x \in {\reali}^4$ will be determined.

Certainly, to determine the $\tilde{\star}$-product, functions $g_{k}$ should
have sufficient smoothness. As stated above the $\star$-product is
well-defined, if $g_{k}$ belongs to one of Gel'fand-Shilov spaces $S^{\beta},
\, \beta < 1/2$ \cite{GSh}. Moreover, $f \star g_{k} \in S^{\beta}$ with the
same $\beta$ \cite{CMTV7}. It gives us the possibility to determine Wightman
functions as generalized functions corresponding to the above mentioned
Gel'fand-Shilov space.

The operator $\varphi_f$ is defined as follows
\begin{equation}\label {vff}
\varphi_f \, g = \{f g_{0}, f \tilde{\star} g_{1}, \ldots f
\tilde{\star} g_{k} \},
\end{equation}
where $f\tilde{\star} g_{i} = f\,(x)\tilde{\star} g_{i}^{1} \,
(x_{1}) \,\tilde{\star} \cdots \tilde{\star}
g_{i}^{i} \, (x_i.)$ \\
We assume that
\begin{equation}\label{br}
f  \, \tilde{\star} \, (g_{i} + h_{i}) = f  \, \tilde{\star} \,
g_{i} + f \, \tilde{\star} \,  {h_{i}}.
\end{equation}
If $\tilde{\star}$   is defined  by one of the equations
(\ref{forallxy}) - (\ref{forxy}), then it is evident that eq.
(\ref{br}) is fulfilled. In accordance with eq. (\ref{br}) any
vector of space $J'_{0}$ is a sum of the vectors belonging to set
$M$, the operator $\varphi_f$ is determined on any vector of space
$J'_{0}$ and $\varphi_f \Phi \in J'_{0}, \forall \; \Phi \in
J'_{0.}$

The scalar product of vectors in $J'_{0}$ we define with the help
of Wightman functions \\ $W\,(x_1,  \ldots, x_n) \equiv \langle
\Psi_0, \varphi\,(x_1)  \, \ldots \, \varphi\,(x_n)  \, \Psi_0
\rangle$. We consider firstly a chain of vectors: vacuum vector
$\Psi_0 = \{1, 0, \ldots 0 \}, \Phi_{1} = \varphi_{f_1} \Psi_0,
\ldots \Phi_{k} = \varphi_{f_k} \ldots \varphi_{f_1} \Psi_0, \;
f_i =
f_i\,(x_i), \, x_i \, \in {\reali}^{4}$. \\
According to (\ref{vff}), $\Phi_{k} = \{0, \ldots f_k
\tilde{\star} \ldots \tilde{\star} f_1, 0 \ldots 0 \}$. \\
Similarly, $\Psi_n = \varphi_{f_{k + 1}} \ldots \varphi_{f_n}
\Psi_0 = \{0, \ldots f_{k+1} \tilde{\star} \ldots \tilde{\star}
f_n, 0 \ldots 0 \}$. It is obvious, that $J'_{0}$ is a span of the
vectors of such a type.

In what follows we consider the case $\tilde{\star} = \star$. Let us point out
that the definition of $ \tilde{\star}$ by formulas (\ref{forallxy}) -
(\ref{forxy}) leads to the final results same with this case.

The scalar product of vectors $\Phi_{k}$ and $\Psi_n$ is
$$
\langle \, \Phi_{k}, \Psi_n \, \rangle = \langle \, \Psi_0,
\varphi_{{\bar{f}}_1} \, \ldots \, \varphi_{{\bar{f}}_k} \,
\varphi_{f_{k + 1}} \, \ldots \, \varphi_{f_n} \, \Psi_0 \,\rangle
=
$$\begin{equation}\label {scprod}
 \int W\,(x_1,  \ldots, x_n) \, \overline {f_{1} \, (x_{1})} \,\star \cdots
\star \,\overline {f_k\,(x_k)} \,\star \,f_{k + 1}\,(x_{k + 1})
\,\star \cdots\star \,f_n\,(x_n) \,d\,x_1 \ldots d\,x_n.
\end{equation}
The adjoined operator ${\varphi}^*_f$ is defined by the standard
formula. If operator $\varphi_f$ is Hermitian then ${\varphi}^*_f
= {\varphi}_{\bar {f}}$. In this paper we consider only Hermitian
(real) operators, but the construction can be easily extended to
complex fields.

Let us show now that a condition
\begin{equation}\label {12}
\langle \, \Phi_{k}, \Psi_n \, \, \rangle = \overline {\langle \,
\Psi_{n}, \Phi_{k} \, \rangle}
\end{equation}
is fulfilled, if (as well as in commutative case \cite{SW}),
\begin{equation}\label {wfe}
W\,(x_1,  \ldots, x_n) = \overline {W\,(x_n,  \ldots,  x_1)}.
\end{equation}
Really, in accordance with (\ref {scprod})
$$\langle \, \Psi_n, \, \Phi_{k} \, \rangle = \langle \, \Psi_0,
\varphi_{\bar{f}_{n}} \, \ldots \, \varphi_{\bar{f}_{k + 1}} \,
\varphi_{f_k} \, \ldots \, \varphi_{f_1} \, \Psi_0 \, \rangle =
$$\begin{equation}\label {skalprod}
\int \, W\,(x_n,  \ldots,  x_1) \, \overline{f_n\,(x_n)}\,\star
\cdots \star\, \overline{f_{k + 1}\,(x_{k +
1})}\,\star\,f_k\,(x_k)\,\star \cdots \star\,{f_1\,(x_1)} \,
d\,x_1 \ldots d\,x_n.
\end{equation}
The required condition is satisfied, since owing to antisymmetry
of $\theta\,^{\mu\nu}$
$$
\overline {f_n\,(x_n)}\star \overline {f_{n-1}\,(x_{n-1})}\star
\ldots\star \overline {f_1\,(x_1)} = \overline {f_1\,(x_1)\star
\ldots\star f_n\,(x_n)}.
$$
According to the formula (\ref {scprod}), the scalar product of
any vectors $g \in M$ and $h \in M$ is
$$
\langle \, g, h \, \rangle = \sum\limits_{k=0}^{\infty}
\sum\limits_{m=0}^{\infty} \, \int \, \, d\,x_1 \ldots d\,x_k \,
d\,y_1 \ldots d\,y_m
$$
\begin{equation}\label {scany}
W\, (x_{1}, \ldots x_k, y_1, \ldots y_m) \, \bar {g_{k}}^{1} \,
(x_{1}) \,\star \,\ldots\star \,\bar {g_{k}}^{k} \, (x_{k})
\,\star \, { h_{m}}^{1} \, (y_{1}) \,\star \, {h_{m}}^{m} \,
(y_{m}).
\end{equation}
As any vector of space $J'_{0}$ is a finite sum of the vectors
belonging to the set $M$, formula (\ref {scany}) defines the
scalar product of any two vectors in $J'_{0}$.

As well as in commutative case, we need to pass from $J'_{0}$ to
nondegenerate and closed space $J$.

The space $J'_{0}$ can contain isotropic, i.e. orthogonal to $J'_{0}$ vectors
which, as is known, form subspace \cite{Bog}. Designating isotropic space as
$\tilde {J_{0}}$ and passing to factor-space $J_{0} = J'_{0}/\tilde{J_{0}}$,
we obtain nondegenerate space, i.e. a space which does not contain isotropic
vectors. Let us note, that if $g \; \in \, \tilde {J_{0}}$ then $\varphi_f g
\; \in \, \tilde {J_{0}}$. For closure of space $J_{0}$ we assume, as well as
in commutative case, that $J_{0}$ is a normalized space. If the metrics of
$J_{0}$ is positive, norm $\Phi \equiv \|\Phi\|$ can be defined by the formula
$\|\Phi\| = {\langle \Phi, \Phi \rangle}^{1/2}$. $\bar {J_{0}}$ is a closure
of $J_{0}$, this closure is carried out with the help of standard procedure -
closure to the introduced norm. In gauge theories in case of covariant gauge
it is necessary to introduce indefinite metric (see e.g. \cite{Strocchi}). Let
us recall that axiomatic formulation was extended on gauge theories, first of
all, in papers of Morchio and Strocchi \cite{MorStr}. The standard space,
which is considered in axiomatic theory of gauge fields, is a Krein space
\cite{Bog}.

Let us recall that nondegenerate space admits fundamental
decomposition if
\begin{equation} \label{dirsum}
J_{0} = J_{0}^{+} \oplus J_{0}^{-},
\end{equation}
where $J_{0}^{\pm}$ is a space with the definite positive
(negative) metric, $J_{0}^{+} \bot \, J_{0}^{-}$. In other words,
if $x \, \in J_{0}$, then  $x = x^{+} + x^{-}, \; x^{\pm} \, \in
J_{0}^{\pm}$, $x^{+} \bot \, x^{-}$, that is ${\langle x^{+},
x^{-} \rangle = 0}.$\\
It is obvious that
\begin{equation} \label{scalxy}
\langle \, x, y \, \rangle = \langle \, x^{+}, y^{+} \, \rangle +
\langle \, x^{-}, y^{-} \, \rangle.
\end{equation}
It is easy to see that we can introduce in $J$ the
positive-definite scalar product $(\cdot, \cdot)$, namely
\begin{equation} \label{scalxymin}
(x, y) = \langle \, x^{+}, y^{+} \, \rangle - \langle \, x^{-},
y^{-} \, \rangle.
\end{equation}
We can introduce norm using this product $\|x\| = {( x,
x)}^{1/2}$. \\ Evidently, $\|x^{+}\| = {\langle \, x^{+}, x^{+} \,
\rangle}^{1/2}, \; \|x^{-}\| = {( - \langle \, x^{-}, x^{-} \,
\rangle )}^{1/2}$. \\
The closure of $J_{0}: \bar{J_{0}} = \bar{J}_{0}^{+} \oplus
\bar{J}_{0}^{-}$ is a Krein space.

The space $\bar{J_{0}}$, in turn, can contain isotropic subspace
$\tilde{J}$. Factor-space $J = \bar {J_{0}}/\tilde {J}$ evidently
is a nondegenerate space.

Thus, we have constructed closed and nondegenerate space $J$ such
that operators $\varphi_f$ are  determined on dense domain
$J_{0}$. Hence, the axiom of cyclicity of vacuum is fulfilled.

Let us construct in $J$ the scalar product of any two vectors
$\Phi$ and $\Psi$. It is obvious, that there exist sequences of
vectors $\Phi^{n} \in J_{0}$ and $\Psi^{m} \in J_{0}$ such that
\begin{equation}\label {fp}
< \Phi, \Psi > = \lim_{n, m \to \infty} < \Phi^{n}, \Psi^{m} >.
\end{equation}
We shall note, that condition $\theta^{0\nu}= 0$ was  not used and
thus given above construction is valid in the general case as
well.

Let us stress that if the $\star$-product acts only in coinciding points and
is substituted by usual one in different points then given construction can
also be fulfilled, only in the different points we have to put
$\theta^{\mu\nu} = 0$. But in this case the function $f\,(x, y) =
f\,(x)f\,(y), x \neq y, f\,(x, x) = f\,(x) \star f\,(x)$ is not continuous in
the points $x = y$. In order to overcome this difficulties let us proceed from
the function defined by eq. (\ref{forxy}) to its regularization:
$$
f_{\alpha}\,(x, y) = \eta\,(x - y)\, \varphi\,(x)  \star
\varphi\,(y),
$$
\begin{equation}\label{reg}
\eta\,(x - y) = 1, \quad \mbox{if} \quad {|x - y|}^{2} < \alpha - \varepsilon,
\quad \eta\,(x - y) = 0, \quad \mbox{if} \quad {|x - y|}^{2} \geq \alpha,
\end{equation}
$\alpha$ is arbitrary, $\varepsilon$ can be taken arbitrary small
without loss of continuity of $\eta\,(x - y)$.

It is evident that if $\alpha \gg \theta$, then expressions
(\ref{formodxy}) and (\ref{reg}) practically coincide.

Now let us pass to the limit $\theta^{\mu\nu} = 0$, that is
proceed to the commutative case. In this case $f\,(x)
\tilde{\star} f\,(y) \rightarrow f\,(x) \,f\,(y)$ and we come to
the construction of space $J$ in the commutative case. Let us
point out that the first step in the  standard construction of
this space \cite{SW}, \cite{Jost} is the introduction of sequences
$g$ determined by the formula (\ref {ser}), in which, however,
$g_{i} \equiv g_{i} \, (x_{1}, \ldots x_{i})$ are smooth functions
of variables $x_{j} \in \, {\reali}^{4}$. We shall note that in
the commutative case, starting with $J'_{0}$, we shall come to the
same space $J$. Really, as space of functions of a type $g_{i}^{1}
\, (x_{1}) \, g_{i}^{2} \, (x_{2}) \ldots g_{i}^{i} \, (x_i)$ is
dense in space of functions $g_{i} \, (x_{1}, \ldots x_{i})$
\cite{SW}, \cite{Jost}, we can complete $J'_{0}$ up to the space
of the above mentioned sequences and then carry out the standard
construction of space $J$.

{ \bf  Remark} \quad{\it In fact we have obtained a very general
construction, which is valid not only in the case of NC QFT, but
also for any case, when usual product of the functions is
substituted by the new one, if the following conditions are
satisfied:

\begin{enumerate}
    \item [i]  Corresponding functions belong to the some space, such that
Wightman functions are defined as generalized functions
(functionals) over this space;

 \item [ii]  The scalar product in this space is defined by eq. (\ref{scprod}), but
condition (\ref{wfe}) may be substituted by the new one;

 \item [iii]  There exists the passage to the standard multiplication.
\end{enumerate}
}

\section{{Conclusions}}
We have rigorously constructed field operators in NC QFT. This
construction is important for any rigorous treatment of the
axiomatic approach to NC QFT via NC Wightman functions and the
derivation of rigorous results such as CPT and spin-statistics
theorems.

The carried out construction of the closed and nondegenerate
space,  such that operators $\varphi_f$ are determined on its
dense domain, corresponds to the theorem named as "main" in
\cite{Jost} and opens a way to derivation of the reconstruction
theorem in noncommutative field theory, that we are going to make.


\begin{thebibliography}{99}

          \bibitem{SW}
 R. F. Streater and A. S. Wightman, {\it PCT, Spin and Statistics and
All That}, Benjamin, NewYork (1964).

          \bibitem{Jost}

2. R. Jost, {\it The General Theory of Quantum Fields },Amer. Math.Soc.,
Providence, R.I. (1965).

          \bibitem{BLOT}
 N. N. Bogoliubov, A. A. Logunov and I. T. Todorov, {\it Introduction to
Axiomatic Quantum Field Theory}, W. A. Benjamin, Inc., New York,
1975; \\
 N. N. Bogoliubov, A. A. Logunov, A. I. Oksak and I. T. Todorov,
{\it General Principles  of Quantum Field Theory}, Kluwer, Dordrecht (1990).

          \bibitem{Haag}
R. Haag, {\it Local Quantum Physics}, Springer, Berlin (1996).


          \bibitem{DN}
M. R. Douglas and N. A. Nekrasov, {\it Rev. Mod. Phys.} {\bf 73} 977 (2001),
hep-th/0106048.

          \bibitem{Sz}
R. J. Szabo,{\it Phys. Rept.} {\bf 378} 207 (2003), hep-th/0109162.

          \bibitem{Snyder}
H. S. Snyder, {\it Phys. Rev.} {\bf 71} 38 (1947).

          \bibitem{Connes}
A. Connes, {\it Noncommutative Geometry},  Academic Press, New York (1994).

          \bibitem{DFR}
S. Doplicher, K. Fredenhagen and J. E. Roberts, {\it Phys. Lett. B} {\bf 331}
39 (1994); {\it Comm. Math. Phys.}, {\bf 172}, 187 (1995).

          \bibitem{SeWi}
N. Seiberg and E. Witten, {\it JHEP} {\bf 9909} 32 (1999), hep-th/9908142.


         \bibitem{AB}
L. \'Alvarez-Gaum\'{e} and J. L. F. Barbon, {\it Int. J. Mod. Phys. A}, {\bf
16} 1123 (2001), hep-th/0006209.

          \bibitem{AGM}
L. \'Alvarez-Gaum\'{e} and M. A. V\'azquez-Mozo, {\it Nucl. Phys. B}, {\bf
668} 293 (2003), hep-th/0305093.

          \bibitem{CMNTV}
M. Chaichian, M. N. Mnatsakanova, K. Nishijima, A. Tureanu and Yu. S. Vernov,
hep-th/0402212.

         \bibitem{VM05}
Yu.S. Vernov, M.N. Mnatsakanova, {\it  Theor. Math. Phys.} {\bf 142} 337
(2005).

          \bibitem{CPT}
M. Chaichian, P. Pre\v snajder and A. Tureanu, {\it Phys. Rev. Lett.}, {\bf
94} 151602 (2005), hep-th/0409096.

        \bibitem{CMTV}
M. Chaichian, M. N. Mnatsakanova, A. Tureanu and Yu. S. Vernov, {\it
 Classical Theorems in Noncommutative Quantum Field Theory}, hep-th/0612112.

          \bibitem{GSh}
I. M. Gelfand and G.E. Shilov, {\it Generalized Functions}, Vol. 2,  Academic
Press Inc., New York (1968).

          \bibitem{CMTV7}
M. Chaichian, M. N. Mnatsakanova, A. Tureanu and Yu. S. Vernov, {\it  JHEP}
{\bf09} 125 (2008), hep-th/0706.1712v1 (2007).

           \bibitem{Sol07}
M. A. Soloviev, {\it Theor. Math. Phys.}  {\bf 153}, 1351 (2007),
math-ph/0708.0811.

          \bibitem{Bog}
J.Bogn\'{a}r, {\it Indefinite Inner Product Spaces}, Springer-Verlag, Berlin,
(1974).

           \bibitem{Strocchi}
F. Strocchi, {\it World Sci.Lect.Notes Phys.}  {\bf 51}, 1 (1993).

           \bibitem{MorStr}
G. Morchio and F. Strocchi, {\it Ann. Inst. H.Poincar\'e A}, {\bf 33}, 251
(1980).
\end{thebibliography}
\end{document}